\documentstyle[twocolumn,epsfig,prb,aps,amsmath,amssymb]{revtex}
\begin{document}

\twocolumn[\hsize\textwidth\columnwidth\hsize\csname
@twocolumnfalse\endcsname

\title{Magnetic order in ferromagnetically coupled spin ladders}

\author{S.~Dalosto and J.~Riera}
\address{
Instituto de F\'{\i}sica Rosario, Consejo Nacional de 
Investigaciones 
Cient\'{\i}ficas y T\'ecnicas, y Departamento de F\'{\i}sica,\\
Universidad Nacional de Rosario, Avenida Pellegrini 250, 
2000-Rosario, Argentina}
\date{\today}
\maketitle
\begin{abstract}
A model of coupled antiferromagnetic spin-1/2 Heisenberg ladders is
studied with numerical techniques. In the case of ferromagnetic
interladder coupling we find that the dynamic and static
structure factor has a peak at $(\pi,\pi/2)$ where the first
(second) direction is along (transversal) to the ladders.
Besides, we suggest that the intensity of this peak and the
spin-spin correlation at the maximum distance along the ladder
direction remain
finite in the bulk limit for strong enough interladder coupling.
We discuss the relevance of these results for magnetic 
compounds containing ladders coupled in a trellis lattice and for
the stripe scenario in high-T$_c$ superconducting cuprates.

\end{abstract}


\vskip2pc]
\section{Introduction}

One of the main topics in condensed matter physics in recent
years has been the study of low-dimensional antiferromagnetic
spin systems.
The strong interest in this field has been sparkled by the
realization that CuO$_2$ planes play an essential role in 
high-T$_c$ superconductors, which was followed by the appearance
of several compounds characterized by the presence of strong
electronic correlations. These compounds include many cuprates,
nickelates, vanadates and manganites, and are characterized by
important and unique properties. In most of them the proximity
of low-dimensional antiferromagnetic (AF) phases are the
key to understand these properties.

At the same time, the concept of spin ladder\cite{ladder0}, originally
introduced to explain the presence of a spin gap in 
(VO)$_2$P$_2$O$_7$ and later in layered cuprates like
Sr$_{n-1}$Cu$_{n+1}$O$_{2n}$ 
(Refs.~\onlinecite{ladd-rice,ladd-rice2}) became an
important theoretical tool to understand the behavior of strongly
correlated systems. The physics of the two-leg spin ladder is
characterized by the existence of a singlet-triplet spin gap
and an exponential decay of correlation functions.
The ground state which can be thought to a good approximation as
a product of singlets living on the rungs is now well understood.
However, the 
above mentioned cuprates, the important case of
Sr$_{14}$Cu$_{24}$O$_{41}$, as well as many other compounds
like CaV$_2$O$_5$, actually contain layers of {\em coupled} 
two-leg ladders. These ladders are coupled by frustrated 
interactions in a trellis lattice which make
its study with analytical or numerical techniques quite
difficult. In principle, in the absence of frustration, a
reduction of the gap as the interladder coupling increases is
expected. Eventually, the system becomes gapless at a quantum
critical point (QCP)\cite{ladd-rice2,qcp} and for larger
coupling it behaves essentially as a two-dimensional (2D) 
spin-1/2 square
antiferromagnet. Much less is known for the trellis lattice,
although a Schwinger boson study\cite{normandmila} suggests
the transition from a spin liquid to a possible spiral order
as the interladder coupling (ILC) increases. Quantum Monte Carlo 
(QMC) studies\cite{miyahara} of this frustrated system are
hampered by the ``minus sign problem" and the possibility
of using this powerful technique is severely reduced. In
this sense, one of the objectives of the present work is to
study a model in which the frustrated AF interladder couplings
of the trellis lattice are replaced by much simpler ferromagnetic
(FM) couplings in a square
lattice. We expect that some of the physics of the frustrated
system can be captured by this effective simplified 
model.\cite{aharony} Besides, there are compounds which consist of
FM coupled ladders like SrCu$_2$O$_3$. The results of the present 
work could be relevant to other FM coupled gaped systems like the
dimerized chains in (VO)$_2$P$_2$O$_7$.\cite{garrett} Previous
studies have compared the behavior of AF and FM frustrated and
nonfrustrated coupled {\em gapless} spin systems 
(spin chains).\cite{whiteaffleck}

Alternatively, a renewed interest in coupled ladders comes
from the high-T$_c$ cuprate superconductors themselves. A
number of recent experiments, mainly neutron scattering
studies\cite{tranquada}, indicate the presence of incommensurate
spin correlations which in turn have been interpreted as coming
from the segregation of charge carriers into 1D domain
walls or ``stripes" leaving the regions between them as
undoped antiferromagnets. There are several theoretical
scenarios that have predicted or that attempt to explain this
stripe order\cite{stripeold,castellani,emery} but the
origin of this order and its relation to superconductivity
is still controversial. In particular, the problem of stripe
formation in a 2D microscopic model like the t-J or Hubbard
models is extremely difficult to study with analytical or
numerical techniques. In principle,
the inclusion of charge and spin degrees of freedom is 
essential for the understanding of this problem. However, it has
been suggested that {\em assuming} the presence of a stripe
structure it is very instructive to study its magnetic
properties by using a model with spin degrees
of freedom only.\cite{tworzydlo,kim} In this simplified
model, the AF insulating regions between the charged stripes
are considered as $n$-leg isotropic ladders coupled by an
effective interaction. In one such study\cite{kim}, following
the initial picture from Ref.~\onlinecite{tranquada}, the
insulating regions were considered as 3-leg ladders coupled
by AF interactions. However, a numerical study of the 2D
t-J model\cite{white}, as well as early studies of charge
inhomogeneities in Hubbard and t-J models\cite{stripeold},
indicate the formation of ``bond-centered" stripes, i.e.
doped two-leg ladders alternating with undoped two-leg 
ladders. Coupled spin two-leg ladders were also 
studied\cite{tworzydlo} but its relevance to the physics of
Cu-O planes is relative since they miss the essential ingredient
that the magnetic order of the AF slices is $\pi$-phase shifted
as emphasized in Ref.~\onlinecite{white}.
Hence, the second motivation for the present work comes from the
assumption that this $\pi$-phase-shift between ladders can be
modeled by taking a {\em ferromagnetic} coupling between them.

In summary, the purpose of the present work is to study 
ferromagnetically coupled two-leg ladders and compare their
behavior with the case of AF coupling. If this model is
considered as an approximation of AF systems in the trellis
lattice, the FM coupling is an effective interaction coming
from the frustrated interactions between ladders. If this
model is considered as an approach to the stripe phase of the
cuprates, the FM interaction comes from a {\em collective} 
effect
determined from the competition of charge and spin degrees
of freedom. In both cases, the results of the present study
lead to predictions which can be tested experimentally. 
We use essentially numerical techniques like QMC (world-line
algorithm) which allows us to reach low enough temperatures
so as to capture ground state properties, and exact
diagonalization with the Lanczos algorithm (LD), complemented by
the continued fraction formalism to compute dynamical properties.

\section{Quasi-one dimensional study.}
\label{coupdim}

\begin{figure} 
\begin{center}
\epsfig{file=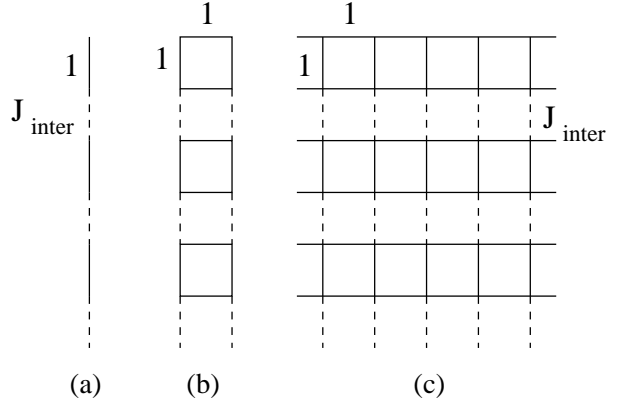,width=8cm,angle=0}
\end{center}
\caption{
(a) Coupled dimers forming an alternating chain; (b) coupled plaquettes
forming an alternating ladder; (c) coupled ladders in a square lattice.
Full lines (dashed) correspond to AF (FM) interactions.
}
\label{figura}
\end{figure}

To gain insight about the effects of FM interladder couplings
we start from the case of FM coupled AF {\em dimers} which are the
simplest systems with a spin gap. We are thus
led to 1D or quasi-1D systems which are much easier to study from
the numerical point of view. Besides, systems with a random
distribution of AF and FM couplings have received some theoretical
attention and their possible physical realization in
SrCuPt$_{1-p}$Ir$_p$O$_6$ has been discussed.\cite{furusaki}
The Hamiltonian is given by:
\begin{eqnarray}
{\cal H}={\cal H}_{dimer} + {\cal H}_{inter}
\nonumber
\end{eqnarray}
where:
\begin{eqnarray}
{\cal H}_{dimer}&=&J \sum_{a}
{\bf S}_{a;1} \cdot {\bf S}_{a;2}, 
\nonumber \\
{\cal H}_{inter}&=& \sum_{a,b,i,j} J_{inter,a,b,i,j}
{\bf S}_{a;i} \cdot {\bf S}_{b;j}, 
\label{hamdim} 
\end{eqnarray}
\noindent
where $a$ is a dimer index and $i=1,2$ labels the sites in a 
dimer. $J=1$ for simplicity. Periodic boundary conditions are
imposed in the longitudinal direction.
There are several ways of coupling dimers. We consider here
the simplest case of dimers forming a FM-AF alternating chain
(Fig.~\ref{figura}(a)).
Another possibility is that of dimers 
forming a two-leg ladder with FM leg and AF rung interactions.
This second case has already been studied numerically\cite{roji} 
but it is not relevant for the problems we wish to address. 
Besides, we consider the case of FM-coupled AF 
{\em plaquettes} instead of dimers in which case we have 
a two-leg ladder with FM-AF alternating interactions along the
legs (Fig.~\ref{figura}(b)). 

\noindent

\begin{figure} 
\begin{center}
\epsfig{file=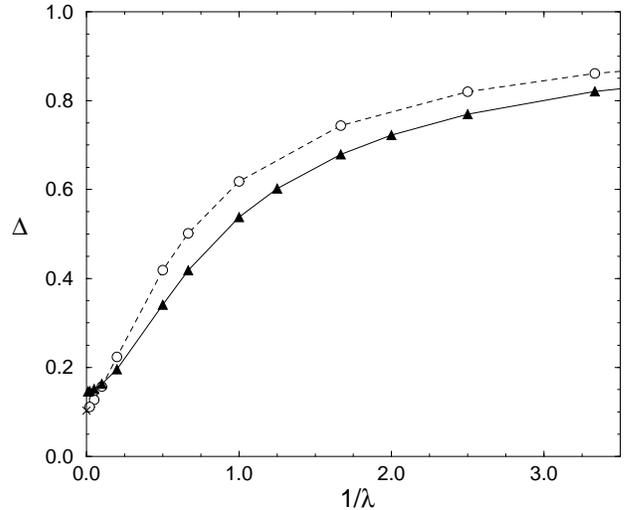,width=7cm,angle=-90}
\end{center}
\caption{
Singlet-triplet spin gap in the bulk limit of FM coupled dimers
(circles) and plaquettes (triangles) as a function of 
$\lambda=-J_{inter}$. The cross indicates the spin gap for the
spin-1 chain $\Delta=0.41$ divided by 4 (from Ref. (19)).
}
\label{onedimgap}
\end{figure}

Our first concern in this section
is the behavior of the spin gap starting from the situation 
of isolated dimers or plaquettes.
Using exact diagonalization we computed the spin gap
$\Delta$ by substracting the 
energies in the $S=0$ and $S=1$ subspaces on finite clusters with
up to 24 sites.
The extrapolation to the bulk limit was done using the law
$\Delta_{\infty} + b \exp(-L/L_0)/L$. The final result is shown in 
Fig.~\ref{onedimgap}.
We notice that in the limit $J_{inter} \rightarrow -\infty$ we

\noindent
recover the cases of a spin-1 chain for the coupled dimer
system and a spin-1 ladder for the coupled plaquette one.
It is easy to realize (by solving a two-dimer system and a two-site
spin-1 system) that the gap for the spin-1 chain is four times
larger than the gap obtained by the coupled dimer system when
$J_{inter} \rightarrow -\infty$ and the gap for the spin-1 ladder is
twice larger than the coupled plaquette system in this limit.
Thus, in the former case we obtain a gap
$\Delta_{cd}\times 4=0.410$, coincident with the value already
reported in the literature.~\cite{whitehuse}
For the spin-1 ladder we would obtain a gap 
$\Delta_{cp}\times 2=0.290 \pm 0.008$, smaller than the gap for the
S=1 chain as predicted theoretically.\cite{allen}
Qualitatively, the important feature here is that the gap decreases
monotonically from the isolated dimers (or plaquettes) case as
$J_{inter} \rightarrow -\infty$.
In the case of FM coupled dimers, a monotonic behavior could be 
guessed from the fact that this system continuously evolves 
towards the valence-bond-solid picture of a spin-1 chain 
in that limit.

\begin{figure} 
\begin{center}
\epsfig{file=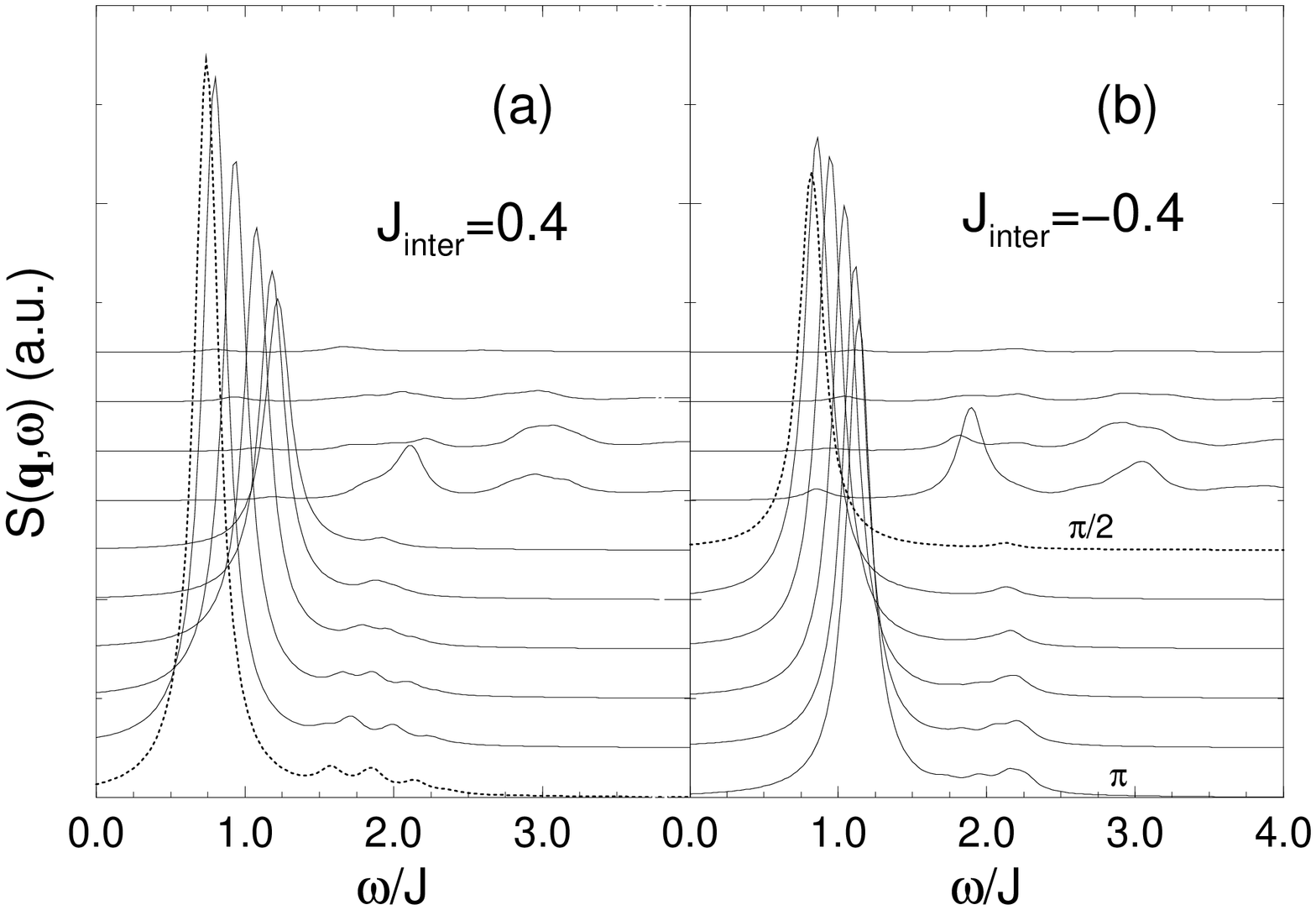,width=6cm,angle=-90}
\epsfig{file=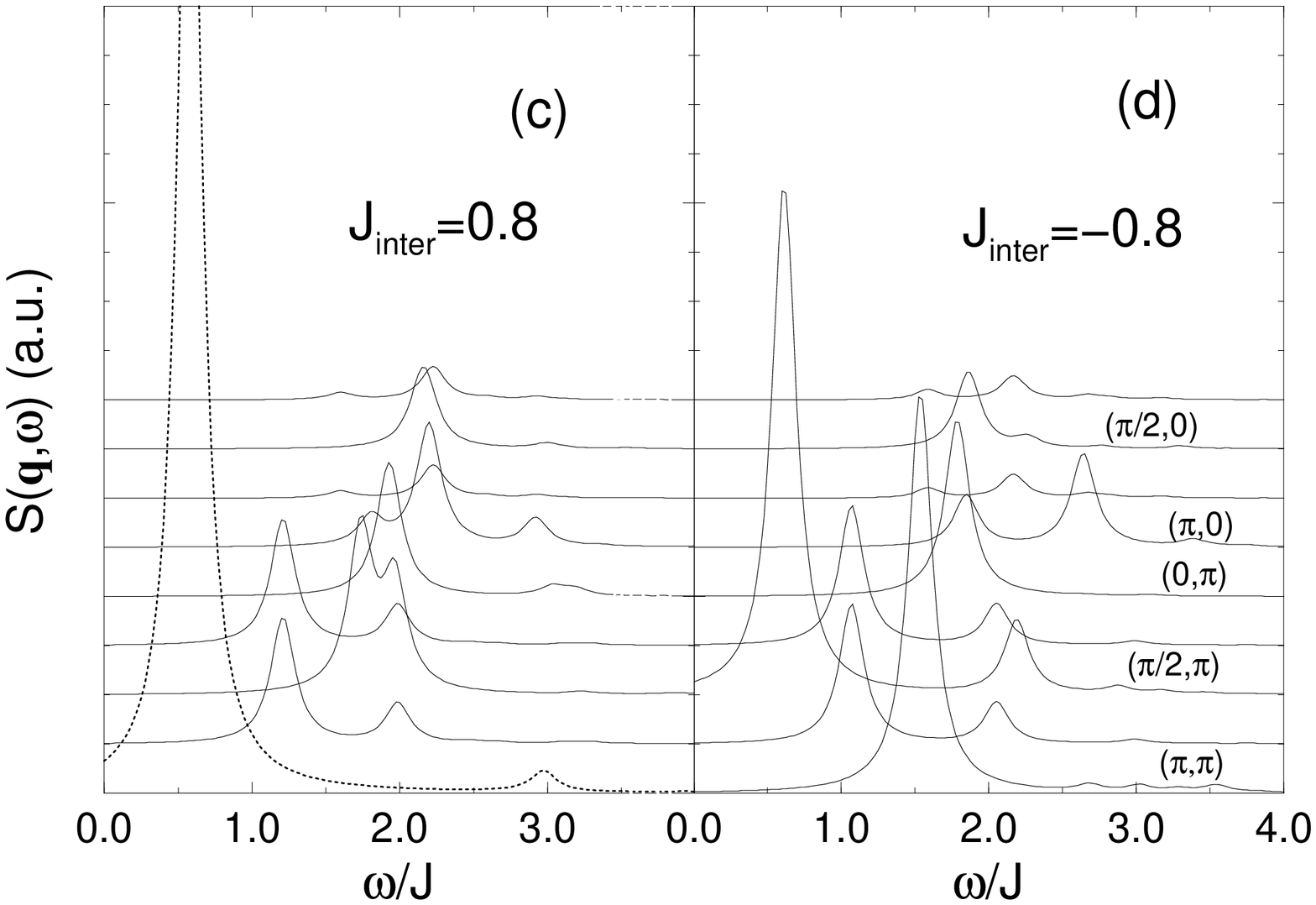,width=6cm,angle=-90}
\end{center}
\caption{
Dynamical structure factor as a function of the frequency for (a) AF
and (b) FM coupled dimers for several momentum increasing from
$q_x=\pi/10$ to $\pi$ in steps of $\pi/10$ from top to bottom.
These results were obtained for the 20 site chain with LD.
The same for (c) AF and (d) FM coupled plaquettes. In this case,
${\bf q}=(q_x,q_y)$, where $x$ is the
longitudinal direction, are shown in steps of $q_x=\pi/4$.
These results were obtained for the 2x8 cluster with LD.
}
\label{dynplaq}
\end{figure}

\noindent
On the other hand, the behavior
of the coupled plaquettes system is not obviously predictable.

The second point we want to examine is the behavior of the excitations
of these systems, in particular the $S=1$ excitations as can be 
measured by neutron scattering experiments. For this purpose, using
conventional Lanczos techniques with the continued fractions formalism,
we have computed the dynamical structure function ($zz$-component)
$S({\bf q},\omega)$\cite{haas} which is shown in Fig~\ref{dynplaq}.
Already for the simplest case of coupled dimers 
(Fig~\ref{dynplaq}(a),(b)) one
can see an interesting feature which we will observe also for the
coupled ladders case in the next section. The position of the gap
which is at $q=\pi$ for AF dimerized chains shifts to $\pi/2$ for the
case of FM coupled dimers.\cite{ariel} A similar behavior is observed
for coupled plaquettes (Fig~\ref{dynplaq}(c),(d)), 
where the lowest energy peak changes from
$(q_x,q_y)=(\pi,\pi)$ to $(\pi/2,\pi)$ ($x$ is the longitudinal
direction) by switching from AF to FM interplaquette couplings.
In both cases, there is a transfer of spectral weight from the
original AF peak to the FM one. This behavior is independent of
the absolute value of $J_{inter}$, except for finite size effects.

\section{Coupled ladders.}
\label{couplad}

We have now arrived at the central part of this work. The 
Hamiltonian for the system we consider now (Fig.~\ref{figura}(c)) 
is essentially the same as (\ref{hamdim})
which we rewrite here for clarity:
\begin{eqnarray}
{\cal H}={\cal H}_{ladder} + {\cal H}_{inter}
\nonumber
\end{eqnarray}
where:
\begin{eqnarray}
{\cal H}_{ladder}&=&J \sum_{a,l,i}
{\bf S}_{a;l,i} \cdot {\bf S}_{a;l,i+1}+ J \sum_{a,i}
{\bf S}_{a;1,i} \cdot {\bf S}_{a;2,i}, 
\nonumber \\
{\cal H}_{inter}&=& J_{inter} \sum_{a,i} 
{\bf S}_{a;2,i} \cdot {\bf S}_{a+1;1,i}, 
\label{hamlad} 
\end{eqnarray}
\noindent
where $a$ stands now for a ladder index and $l=1,2$ for the
two legs in a ladder. We have taken the case of an isotropic 
ladder by simplicity. $J$ is again taken as the unit of energy.

\begin{figure}
\begin{center}
\epsfig{file=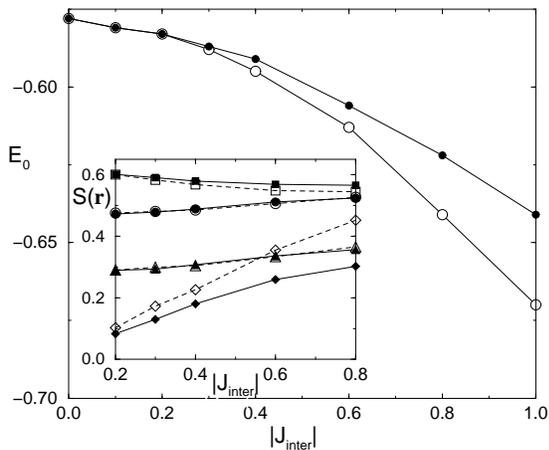,width=6cm,angle=-90}
\end{center}
\caption{Ground state energy per site in the bulk limit
of AF (open circles) and
FM (full circles) coupled ladders as a function of the absolute
value of $J_{inter}$. Error bars are smaller than the symbol
size. In the inset, the intraladder (1,0) (circles), (0,1) 
(squares), (1,1) (triangles), and interladder (0,1) (diamonds) 
spin-spin correlations are shown as a function of $|J_{inter}|$
for the $16 \times 16$ cluster.
Open (full) symbols correspond to AF (FM) interladder couplings.
}
\label{enerlad}
\end{figure}

We start with the study of the ground state energy of the 
system of both FM and AF coupled ladders. To this purpose
we have performed standard QMC simulations for $L \times L$ 
lattices with $L=4,6,8,12,$ and 16.
Periodic boundary conditions in both directions are considered.
For each lattice and set of coupling constants, we 
took ${\rm T}/J= 0.125,~0.100$ and 0.07, and the Trotter number 
$M$ at each temperature such that the error due to the time
discretization is comparable or smaller than the statistical
error. Typically, $M=140$ for ${\rm T}/J= 0.07$. We made runs up
to $10^6$ MC steps for both thermalization and measurement. 
We computed the energy in the total $S_z=0$ and $S_z=1$ 
subspaces ($E_0$ and $E_1$ respectively). For each set of 
coupling constants, we extrapolated the corresponding energies
per site $e_0$ and $e_1$ to the bulk limit using the law 
$e_{\infty} + b \exp(-L/L_0)/L^2$ in the gaped region and
$e_{\infty} + b /L^3$ in the gapless case. 
We obtained very close values
for $e_{\infty ,0}$ and $e_{\infty ,1}$ which gives an indication
of the good quality of the fits.
The results for the ground state energy for
FM and AF ladder couplings are shown in Fig.~\ref{enerlad}. An
interesting feature can be noticed: the energies for both signs
of $J_{inter}$ are degenerate within numerical errors for
$|J_{inter}| < 0.3$. This situation corresponds to a
physics governed mainly by singlet dimers on the ladder rungs
and in this case
the sign of the coupling between these relatively isolated dimers
is irrelevant. On the other hand, for $|J_{inter}| > 0.3$, for
AF interladder coupling it is possible that the singlets
delocalize from a single ladder and finally form a ``resonant
valence bond" state or that the singlet-triplet
excitations be replaced by gapless magnon excitations. 
Previous numerical studies\cite{imada,tworzydlo} precisely
locate at $J_{inter} \approx 0.3$ the position of the QCP at which the
ladder-like spin liquid is replaced by a long range 2D-like
AF order thus choosing the second possibility.
The important point we want to suggest is that in both
cases the energy of the system would be lower than for the
case of FM ILC where the singlets on the ladders still
persist. To illustrate this scenario we have computed on the 
$16 \times 16$ cluster the spin-spin correlation functions
$S({\bf r})=\langle S^z_0 S^z_{\bf r} \rangle$
for ${\bf r}=(1,0)$ (leg direction), (0,1) (rung), and (1,1),
inside a ladder, and ${\bf r}=(0,1)$ between
two ladders. These correlations, normalized in such a way that 
$S(0)=1$, are shown in absolute value
in the inset of Fig.~\ref{enerlad} as a function of $|J_{inter}|$.
The differences between
FM and AF ILC appear in the (0,1) (rung) correlations, which
remain stronger in the former case, and most importantly, in the 
interladder

\noindent
(0,1) correlation which increases faster in the latter. Of
course, this correlation is negative (positive) in the
AF (FM) case.

\begin{figure}
\begin{center}
\epsfig{file=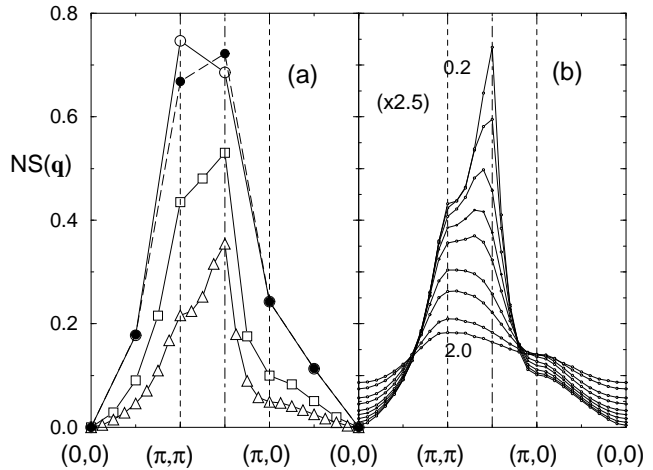,width=5.5cm,angle=-90}
\end{center}
\vspace{0.4cm}
\caption{
Static structure factor obtained by QMC for $J_{inter}=-0.2$.
(a) For various cluster sizes. The clusters
considered are $4\times 4$ (circles), $8\times 8$ (squares),
and $16\times 16$ (triangles). The curve with full circles
corresponds to the $4\times 4$ cluster and $J_{inter}=-0.3$.
(b) For the $20 \times 20$ cluster at $\rm T=0.2, 0.3, 0.4, 0.5,
0.6, 0.8, 1.0, 1.5$ and 2.0. The curves have been multiplied
by 2.5 for clarity.
}
\label{static}
\end{figure}

This indication of a difference between the two ILC cases can be 
traced to a more intimate level which would also
provide experimentally measurable features. To this end, let us
examine now the static structure factor $S({\bf q})$ obtained
by Fourier transforming the spin-spin correlations obtained
by QMC at the lowest temperature attained.
In the case of AF ILC the peak is at $(\pi,\pi)$ in all the
range from the isolated ladder, which corresponds to the gaped 
``quantum disordered" region, to the isotropic square
lattice, but the extrapolation of its intensity to the
bulk limit becomes nonzero only for 
$J_{inter} > J_{inter,cr}$, i.e. in the ``renormalized classical"
region.\cite{kim,tworzydlo}
For FM ILC, as shown in Fig.~\ref{static}(a)
for $J_{inter}=-0.2$, the situation is qualitatively different.
The peak in $S({\bf q})$ is now at $(\pi,\pi/2)$, a feature which
is similar to the one seen in the simpler cases of 
coupled chains and plaquettes.
This behavior has been found for all clusters considered, and for
all $J_{inter} <0$ except for finite size effects: the smaller
$|J_{inter}|$ the larger the size needed to reach the bulk behavior.
This is illustrated for the $4\times 4$ cluster.

A highly nontrivial
behavior is found if the temperature dependence is analyzed. In 
Fig.~\ref{static}(b)) the evolution of the structure factor for the
$20 \times 20$ cluster and $J_{inter}=-0.2$ is shown at $\rm T=0.2, 
0.3, 0.4, 0.5, 0.6, 0.8, 1.0, 1.5$ and 2.0. In the high temperature 
region ($\rm T>0.8$) the peak is located at $(\pi,\pi)$. As T is 
lowered the peak starts to shift towards the zero temperature peak 
$(\pi,\pi/2)$ which is reached at $\rm T=0.3$. We found almost no 
variation with cluster size of these two crossover temperatures at
this value of $J_{inter}$. The fact that only at a finite temperature
the peak of the magnetic structure factor starts to be incommensurate 
is reminiscent to the one first found in 
La$_{1.6-x}$Nd$_{0.4}$Sr$_x$CuO$_4$ ($x \approx 0.8$) where a 
charge-stripe order is developed at $\rm T_c=65 K$ followed by a 
spin-stripe order at a {\em lower} temperature 
$\rm T_s=50 K$.\cite{tranquada}

\begin{figure}
\begin{center}
\epsfig{file=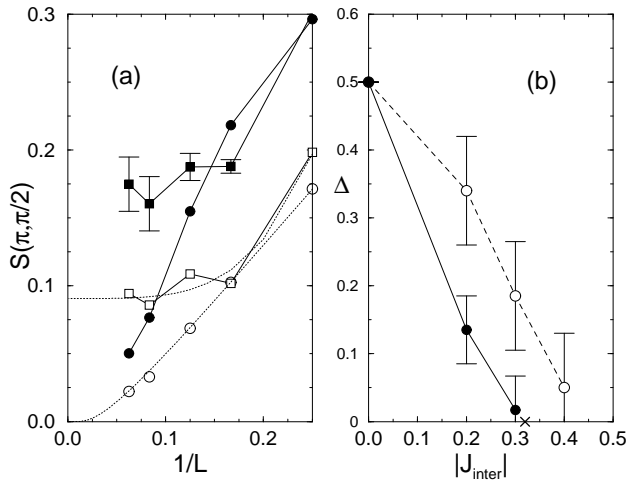,width=6.5cm,angle=-90}
\end{center}
\caption{
(a) $S(\pi,\pi/2)$ (open symbols) and $S_{max,x}$ (full symbols)
as a function of $1/L$ for $J_{inter}=-0.2$ (circles) and $-0.6$
(squares).
(b) Singlet-triplet spin gap in the bulk limit of FM (open circles) and
AF (full circles) coupled ladders as a function of the absolute
value of the interladder coupling constant. In the AF case, the
value of $J_{inter}$
at which $\Delta=0$ is taken from Ref.~23.
The lines are guides to the eye.
}
\label{cmaxlad}
\end{figure}

An important quantity to compute is the bulk limit of the
peak of the structure factor. This quantity is related with
the behavior of spin-spin correlations at
the maximum distance along the ladder direction ($x$-axis), 
along the direction transversal to the ladders ($y$-axis) and 
at the maximum distance of the 2D cluster.
In the case of the AF square lattice, this latter quantity
is proportional in the bulk limit to the squared staggered 
magnetization and it should be equal in that limit to the
static structure factor at momentum $(\pi,\pi)$.\cite{reger}
The finite size scaling of $S(\pi,\pi/2)$ is shown in
Fig.~\ref{cmaxlad}(a) for $J_{inter}=-0.2$ and $-0.6$.
We have attempted extrapolations to the bulk limit using both
exponential and power laws.
Due to the fact that clusters with $L=4,8,16$ and $L=6,12$
belong to two different sets (which is
more noticeable for large values of $|J_{inter}|$),
the extrapolation procedure is not very reliable.
However, as shown in Fig~\ref{cmaxlad}(a), one can 
conclude that $S(\pi,\pi/2)$ is zero for 
$J_{inter}=-0.2$ and nonzero for $J_{inter}=-0.6$.
The finite size behavior of the spin-spin correlation at the
maximum distance along the ladder direction, $S_{max,x}$, which
is the one with smaller errors in our simulations, is
similar to the one for $S(\pi,\pi/2)$ and the extrapolated
values are also zero (nonzero) for $J_{inter}=-0.2$ 
($J_{inter}=- 0.6$).

This crossover in the behavior of $S(\pi,\pi/2)$ as a function of
$J_{inter}$ poses us with the question of the existence of a
point analogous to the QCP in the AF ILC case. In the limit of
$J_{inter}\rightarrow -\infty$ the coupled ladder system becomes
equivalent to a system of AF coupled spin-1 chains, where a finite
coupling is necessary to change to a gapless regime.\cite{azzouz}
To answer this question, let us now examine the behavior of the
singlet-triplet spin gap as a function of $J_{inter}$.
Although this is not a convenient quantity to compute with QMC since it
involves a difference between absolute values of the energies
and then for large clusters the error becomes comparable to its
value, we could get an indication of the presence or absence of
a gapless region.
The gap was computed for finite clusters and
then extrapolated to the bulk using the law 
$\Delta_{\infty} + b \exp(-L/L_0)/L$ (or $ a /L^2$ for the
gapless case\cite{manousakis}). The results are depicted
in Fig.~\ref{cmaxlad}(b). For the AF case, it can be seen a rapid 
decrease of $\Delta$ as $J_{inter}$ is increased, confirming
earlier predictions and 
calculations.\cite{ladd-rice2,imada,tworzydlo}
The gap vanishes at the QCP, $J_{inter,cr} \approx 0.3$.
For FM ILC we also obtain a monotonically decreasing
behavior, similar to the one found in the previous section
for coupled dimers and plaquettes. The gap seems larger to that
of AF ILC but it could vanish at $J_{inter}\approx -0.4$ within error
bars. The calculation of other quantities like correlation lengths
and/or using more powerful techniques should be necessary to 
obtain a reliable estimation for $J_{inter,cr}$.

\begin{figure}
\begin{center}
\epsfig{file=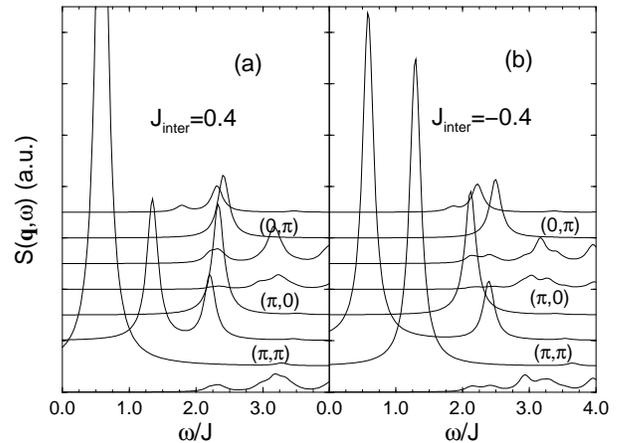,width=6cm,angle=-90}
\end{center}
\caption{
Dynamical structure factor obtained by LD on the $4\times 4$
cluster vs. frequency at several momenta
${\bf q}$ for (a) AF and (b) FM interladder couplings.
From bottom to top the values of
${\bf q}$ are $(\pi/2,\pi/2)$, $(\pi,\pi)$, $(\pi,\pi/2)$, $(\pi,0)$,
$(\pi/2,0)$, $(\pi/2,\pi)$, $(0,\pi)$ and $(0,\pi/2)$.
}
\label{dynamic}
\end{figure}

The final part of our study which can eventually lead to a
deeper understanding of the excitations involved in this
system is the analysis of the dynamical structure factor
$S({\bf q},\omega)$ which has been done with LD as in the 
previous section. In this case, we have to limit ourselves to 
somewhat smaller clusters but we hope that the qualitative 
features we found will survive in the bulk limit. Results
obtained for the $4 \times 4$ cluster are shown in 
Fig.~\ref{dynamic}. For AF ILC (Fig.~\ref{dynamic}(a)) the peak
in $S({\bf q},\omega)$ is located at $(\pi,\pi)$, as expected
in the bulk limit for an AF order. When a FM ILC is involved
(Fig.~\ref{dynamic}(b)) it can be seen that considerable 
spectral weight is transferred to the peak at $(\pi,\pi/2)$,
which becomes also the lowest energy excitation.
Results for the $6 \times 4$ cluster are quite similar and it
is quite reassuring that these results are consistent with
the ones obtained with QMC and shown in Fig.~\ref{static}.

\section{Conclusions}

In summary, we have numerically obtained some exact (except
for extrapolation procedures) results for ferromagnetically
coupled systems, in particular two-leg ladders. 
Our main results are embodied in Fig.~\ref{static}
and Fig.~\ref{dynamic}, i.e. at zero temperature the peak of the 
structure factor is
located at $(\pi,\pi/2)$ and it corresponds to the lowest
energy excitation. Fig.~\ref{cmaxlad}(a) also suggests finite
values of this peak of the structure factor and the spin-spin
correlation at the maximum distance
along the ladder axis in the bulk limit for strong $J_{inter}$.
There also two crossover temperatures, a higher one at which the peak
starts to shift away from $(\pi,\pi)$ and a lower one at which
the peak reaches its zero temperature position.
Besides the intrinsic interest for the theoretical understanding
of spin-1/2 ladder
systems, we will try to emphasize in this section their
possible relevance for realistic compounds containing ladders.
As mentioned in the introduction, we should consider in the
first place ladder compounds like SrCu$_2$O$_3$\cite{azuma},
Sr$_{14}$Cu$_{24}$O$_{41}$ (which upon Ca-doping and under
pressure becomes superconducting\cite{uehara}) and 
CaV$_2$O$_5$. In these compounds, the ladders are coupled
forming a trellis lattice. In the former case the interladder
couplings are actually ferromagnetic.
In the others, due to the
frustrated nature of the AF ILC one could speculate that to
some extent they could be modeled effectively by FM couplings.
For all these compounds, then we predict that neutron
scattering experiments would show peaks at 
$(q_x,q_y)=(\pi,\pi/2)$, where the $x$ ($y$) axis is in the
direction parallel (perpendicular) to the ladders.

As also suggested in the introduction, our results could be
related to the striped structure which dynamically appears in
the Cu-O planes of high-T$_c$ cuprates. In this case we are
able to trace the origin of the neutron scattering peaks
observed away from $(\pi,\pi)$ to an effective ferromagnetic
interaction between $\pi$-shifted insulating spin ladders.
In fact, this behavior can be observed already for the simplest
case of ferromagnetically coupled AF dimers as shown in Section
\ref{coupdim}. In this case, it is easy to verify on small
chains by LD or QMC that the peak moves continuously from
$q=\pi/2$ to $\pi$ as some of the FM couplings are replaced
by AF ones. If this picture could translate to coupled ladders,
then one would be lead to the conclusion that some of the
spin two-leg ladders are $\pi$-shifted while others are in phase
in order to reproduce the experimentally observed incommensurate
peaks. Another feature we want to emphasize is the temperature
evolution of the structure factor (Fig.~\ref{static}(b)): the
peak at $(\pi,\pi/2)$ is reached at a {\em finite} temperature
and there is a range of temperatures in which an incommensurate
peak is present. As mentioned in the previous section, this is
reminiscent of the order in which charge and the spin stripes
appear in the cuprates as the temperature is decreased.
The fact that the spin gap possibly remains finite for
somewhat strong values of $|J_{inter}|$ is also interesting for 
the stripe scenario of cuprates although in this case our results
for the bulk limit are affected by large
error bars. Of course, the question of to what extent this model
of FM coupled ladders could apply to this scenario should come
of detailed comparison with more realistic models like 2D
t-J or Hubbard models.

\acknowledgements

We wish to acknowledge many interesting discussions with A. Dobry,
A. Greco and A. Trumper.
We thank the Supercomputer Computations Research Institute (SCRI)
and the Academic Computing and Network Services at Tallahassee
(Florida) for allowing us to use their computing facilities.


\begin{references}

\bibitem{ladder0} E.~Dagotto, J. Riera, and D. Scalapino, Phys. 
       Rev. B {\bf 45}, 5744 (1992); T. Barnes, E.~Dagotto, J.
       Riera, and E. Swanson, Phys. Rev. B {\bf 47}, 3196 (1993).
       For a recent review see, E.~Dagotto, cond-mat/9908250.

\bibitem{ladd-rice} T. M. Rice, S. Gopalan, and M. Sigrist,
      Europhys. Lett. {\bf 23}, 445 (1993).

\bibitem{ladd-rice2} S. Gopalan, T. M. Rice, and M. Sigrist,
      Phys. Rev. B {\bf 49}, 8901 (1994).

\bibitem{qcp} S. Chakravarty, B. I. Halperin, and D. R. Nelson,
      Phys. Rev. B {\bf 39}, 2344 (1989); A. V. Chubukov, S. Sachdev,
      and J. Ye, Phys. Rev. B {\bf 49}, 11919 (1994).
 
\bibitem{normandmila} B.~Normand, K. Penc, M. Albrecht, and 
     F.~Mila, Phys. Rev. B {\bf 56}, 5736 (1997).

\bibitem{miyahara} S. Miyahara, M. Troyer, D. C. Johnston,
      and K. Ueda, J. Phys. Soc. Jpn. {\bf 67}, 3918 (1998).

\bibitem{aharony} This relation between ferromagnetic and
      frustrated interactions has been often exploited in the
      literature. See eg., A. Aharony, R. J. Birgeneau, A.
      Coniglio, M. A. Kastner, and H. E. Stanley, Phys.~Rev.
      Lett. {\bf 60}, 1330 (1988).

\bibitem{garrett} A. W. Garrett, S. E. Nagler, D. A. Tennant, B.
   C. Sales, and T. Barnes, Phys.~Rev.~Lett. {\bf 73}, 2626 (1994)

\bibitem{whiteaffleck} S. R. White and I. Affleck, Phys.~Rev.~B 
       {\bf 54}, 9862 (1996).

\bibitem{tranquada} J. M. Tranquada {\it et al.}, 
     Nature {\bf 375}, 561 (1995); Phys. Rev. B {\bf 54}, 7489
     (1996); K. Yamada {\it et al.}, Phys.~Rev.~B {\bf 57}, 
     6165 (1998); N. Ichikawa
     {\it et al.}, cond-mat/9910037, and references therein.

\bibitem{stripeold} J. Zaanen and O. Gunnarson, Phys.~Rev.~B 
      {\bf 40}, 7391 (1989); D. Poilblanc and T. M. Rice,
      Phys.~Rev.~B {\bf 39}, 9749 (1989); U. L\"ow, V. J.
      Emery, K. Fabricius, and S. A. Kivelson, Phys.~Rev.
      Lett. {\bf 72}, 1918 (1994);
      S. Haas, E. Dagotto, A. Nazarenko, and J. Riera,
      Phys.~Rev.~B {\bf 53}, 8848 (1996).

\bibitem{castellani} C. Castellani, C. Di Castro, and M. Grilli,
     Phys.~Rev.~Lett. {\bf 75}, 4650 (1995).  

\bibitem{emery} V. J. Emery, S. A. Kivelson, and O. Zachar,
     Phys.~Rev.~B {\bf 56}, 6120 (1997), and references therein.

\bibitem{tworzydlo} J. Tworzydlo, O. Y. Osman, C. N. A. van Duin,
     and J. Zaanen, Phys. Rev. B {\bf 59}, 115 (1999).

\bibitem{kim} Y. J. Kim {\it et al.}, cond-mat/9902248.

\bibitem{white} S. R. White and D. J. Scalapino, 
     cond-mat/9705128; cond-mat/9907375.

\bibitem{furusaki} A.~Furusaki, M.~Sigrist, P. A. Lee, K. Tanaka,
     and N. Nagaosa, Phys.~Rev.~Lett. {\bf 73}, 2622 (1994).

\bibitem{roji} M.~Roji and S.~Miyashita,
            J. Phys. Soc. Jpn. {\bf 65}, 883 (1996).

\bibitem{whitehuse} S. R. White and D. Huse, Phys.~Rev.~B {\bf 48},
                 3844 (1993).

\bibitem{allen} D. Allen and D. S\'en\'echal, cond-mat/9908224.

\bibitem{haas} See e.g., S.~ Haas, J.~Riera, and E.~Dagotto,
         Phys.~Rev.~B {\bf 48}, 3281 (1993).

\bibitem{ariel} A first order perturbation expansion around isolated
       dimers is enough to understand this behavior (A. Dobry, private 
       communication).

\bibitem{imada} M.~Imada and Y.~Iino, J. Phys. Soc. Jpn.
      {\bf 66}, 568 (1997).

\bibitem{azzouz} M. Azzouz and B. Doucot, Phys. Rev. B {\bf 47},
      8660 (1993).

\bibitem{manousakis} E. Manousakis, Rev. Mod. Phys. {\bf 63},
      1 (1991), and references therein.

\bibitem{reger} J. Reger and A. P. Young, Phys. Rev. B {\bf 37},
      5978 (1988).

\bibitem{azuma} M. Azuma {\it et al.}, Phys. Rev. Lett. {\bf 73},
     3463 (1994).

\bibitem{uehara}M. Uehara, T. Nagata, J. Akimitsu, N. Mori, and
      K.Kinoshita, J. Phys. Soc. Jpn. {\bf 65}, 2764 (1996).

\end{references}
\end{document}